# Multi-principal element alloy discovery using directed energy deposition and machine learning

Phalgun Nelaturu[1], Jason R. Hattrick-Simpers[2], Michael Moorehead[3], Vrishank Jambur[1], Izabela Szlufarska[1], Adrien Couet[3], Dan J. Thoma[1*]

[1]Department of Materials Science and Engineering, University of Wisconsin, Madison, WI, USA
[2]Department of Materials Science and Engineering, University of Toronto, ON, Canada
[3]Department of Engineering Physics, University of Wisconsin, Madison, WI, USA

*Corresponding authors: dthoma@wisc.edu

**Abstract**

Multi-principal element alloys open large composition spaces for alloy development. The large compositional space necessitates rapid synthesis and characterization to identify promising materials, as well as predictive strategies for alloy design. Additive manufacturing via directed energy deposition is demonstrated as a high-throughput technique for synthesizing alloys in the Cr-Fe-Mn-Ni quaternary system. More than 100 compositions are synthesized in a week, exploring a broad range of compositional space. Uniform compositional control to within ±5 at% is achievable. The rapid synthesis is combined with conjoint sample heat treatment (25 samples vs 1 sample), and automated characterization including X-ray diffraction, energy-dispersive X-ray spectroscopy, and nano-hardness measurements. The datasets of measured properties are then used for a predictive strengthening model using an active machine learning algorithm that balances exploitation and exploration. A learned parameter that represents lattice distortion is trained using the alloy compositions. This combination of rapid synthesis, characterization, and active learning model results in new alloys that are significantly stronger than previous investigated alloys.

## 1. INTRODUCTION

Material development has historically demonstrated incremental advances in properties with modest compositional additions to a base alloy [1]. Examples of primary constituent alloys that have evolved with micro-alloy additions focused on specific properties for an optimized structural application include Al [2], Ni [3-5], Fe [6,7], and Zr [8,9] alloys. To establish a new paradigm for materials discovery and implementation, possibly at higher discovery and implementation rates, high entropy alloys (HEAs) or multi-principal element alloys (MPEAs) offer the potential for new material properties that capitalize on the exploration of combining multiple alloying elements [10,11]. As a specific example, the Cantor alloy [12,13], equimolar Co-Cr-Fe-Mn-Ni, and its various derivatives have been evaluated for enhanced properties such as fracture toughness [14,15], corrosion resistance [16-18], high-temperature strength [19,20], fatigue strength [21,22], and irradiation resistance [23-26]. Most studies have focused on equimolar compositions of alloys seeking to take advantage of the proposed high configurational entropy [27,28]. However, deviating from equimolar compositions opens a vast unexplored composition space that could hold promising properties for a wide range of applications. To explore a potential composition space of nearly a million alloys within a five-component alloy system [10,12,13,29], high-throughput strategies are needed to synthesize and evaluate these materials.

One of the popular, state-of-the-art routes of high-throughput materials synthesis has been combinatorial thin film libraries [30-32]. Large compositional spaces can be rapidly explored with this technique. However, drawbacks include a lack of bulk samples that can be subjected to macroscale testing because the samples are confined to thin films. Furthermore, with extreme cooling rates of up to $10^{10}$ K/s, the thin film samples exhibit differing metastable equilibrium conditions compared to bulk samples [33,34]. Other promising high-throughput techniques include additive manufacturing of compositional gradients [35-38] and diffusion multiples [39-41]. All these methods have a common limitation – a lack of bulk samples with a single alloy composition. The limited sample size restricts the ability to characterize, test, and evaluate complete sets of material properties. To overcome these limitations, we previously proposed a strategy to fabricate bulk samples of different compositions using in situ alloying during directed energy deposition (DED) [42]. Albeit high-throughput, this prior methodology did not include any method for materials design or prediction.

To fully exploit high-throughput methods of alloy fabrication and characterization of HEAs/MPEAs, new design strategies are necessary. For example, machine learning (ML) techniques may be employed, which use mathematical methods to predict trends from datasets that do not necessarily have the corresponding physics-based models. Currently, most supervised ML models use data from previous simulations, experimental results from the literature, or existing databases to predict desired responses [43-46]. The researcher then featurizes their material and trains a model that uses those features to predict the desired property or properties. The generalizability of the model is often confirmed by employing cross-validation so that predictions are made on data the trained model has not yet been exposed to. However, the lack of expansive datasets can limit the design of multi-component alloy systems. For example, constrained thermodynamic datasets for compositional phase equilibria coupled with incomplete physical property assessments hinder a comprehensive ability to computationally design new materials.

In contrast, this research uses a batch active learning framework – our DED-enabled rapid synthesis technique allows the generation of raw data that can be used to train ML models. Initial data are used to train an ML model which is used to predict the properties of new alloys. We then synthesize those alloys to test and validate the model predictions. There is no reliance on external databases for the raw data of alloy compositions or properties, only for the material descriptors of the synthesized alloys and locally made measurements. We believe that this type of closed-loop prediction is the basis for developing autonomous alloy composition design strategies. Specifically, this work demonstrates high-throughput alloy development by combining high-throughput DED synthesis and high-throughput characterization with an active learning-based ML algorithm.

## 2. METHODS

### 2.1. In situ alloying via DED

High-throughput alloy synthesis was achieved by DED-based additive manufacturing in an Optomec LENS MR-7 system that uses a 1-kW Nd:YAG 1070-nm-wavelength laser with a 600-μm spot size. The unit is equipped with four powder hoppers that are independently controlled by augers located at their base. The rotation of these augers allows the powder from the hoppers to enter a line of pressurized Ar gas that carries them into the path of the laser to be melted and deposited on a build plate. The quantity of powder coming out of each hopper can be controlled by the revolutions per minute (RPM) of the respective auger. A schematic of the process is shown in Figure 1.

Each powder hopper was filled with a single elemental powder: Cr, Fe, Mn, or Ni. Having a single elemental powder in each of the four powder hoppers enabled the independent control of the amount of an individual element flowing into the laser path and, thereby, the final composition of the built sample. The powders were procured from American Elements in gas-atomized form with a size distribution of ~45-150 μm (shown in Figure 1). Cr and Ni powder particles were nearly completely spherical, although some of the Ni powder exhibited tiny satellite particles sticking to larger particles. Fe and Mn powder particles were irregular in shape. Despite the morphology differences, all the powders flowed well and exhibited a linear relationship of increasing mass flow rates with increasing auger RPMs. The details of the powder mass flow calibrations are provided in the Supplementary section.

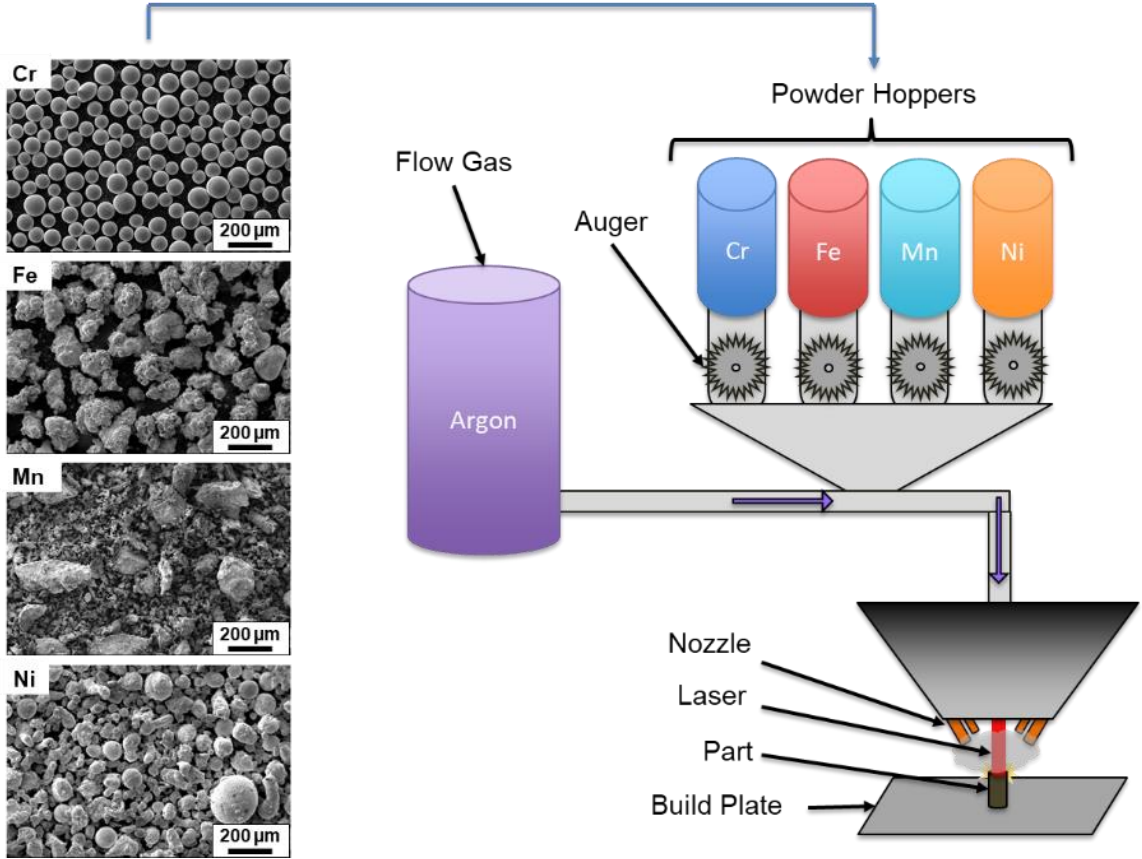

**Figure 1:** *Elemental powders used in this work (left). Schematic of the in situ alloying process (right) reproduced from [42].*

The printing was carried out maintaining an inert Ar atmosphere inside the main chamber at a slight positive pressure (+0.003 atm) above the ambient atmospheric pressure and keeping the oxygen level below 5 ppm. Prior to printing actual alloys, it was necessary to optimize the laser parameters for print and remelt layers, as well as calibrate the mass flow rates of the powders to obtain target compositions in the final alloys.

First, the ideal printing parameters were determined by building 25 samples in a 5×5 array on a build plate, varying the laser parameters but maintaining the predicted composition constant at an equimolar ratio. The laser powers were varied from 300 to 500 W in increments of 50 W, and the laser scan speeds were varied from 10 to 30 in/min (4.2 to 12.7 mm/s) in increments of 5 in/min (2.1 mm/s). A single remelt pass was performed after each print layer with a laser power of 400 W and scan speed of 50 in/min (21.2 mm/s). The hatch spacing was 0.381 mm for the print layers and 0.19 mm for the remelting layers. A laser power of 350 W and scan speed of 25 in/min (10.6 mm/s) were selected as the optimum print parameters based on which sample built closest to the target dimensions – 6.35×6.35 mm and height 5 mm.

Next, the remelt parameters were optimized by similarly building 25 equimolar samples on a build plate. The laser powers varied from 200 to 600 W in increments of 100 W, and the number of remelt passes between each print layer was varied from 1 to 5. The laser scan speed was maintained at 50 in/min (21.2 mm/s). The printing parameters for each print layer were 350 W at 25 in/min (10.6 mm/s), based on the previous plate. The hatch spacing was 0.381 mm for the print layers and 0.19 mm for the remelting layers. The samples were then leveled to the same height, polished, and large-area EDS scans were performed to quantify the area fraction of the unmelted powders. The results are shown in Figure 2 which plots the area fraction of unmelted Cr against the laser power for one, three, and five remelt passes. A clear trend emerged – the area fraction of the unmelted Cr decreased rapidly with increasing laser power, dropping to 0.1% or less at 600 W. The number of remelt passes had a lower effect on the melting of the powders, but in general, the area fraction of unmelted powders decreased with increasing number of passes. Three remelt passes at a laser power of 500 W were selected as the optimum parameters that exhibited the least fraction of unmelted powders while still maintaining close-to-target dimensions.

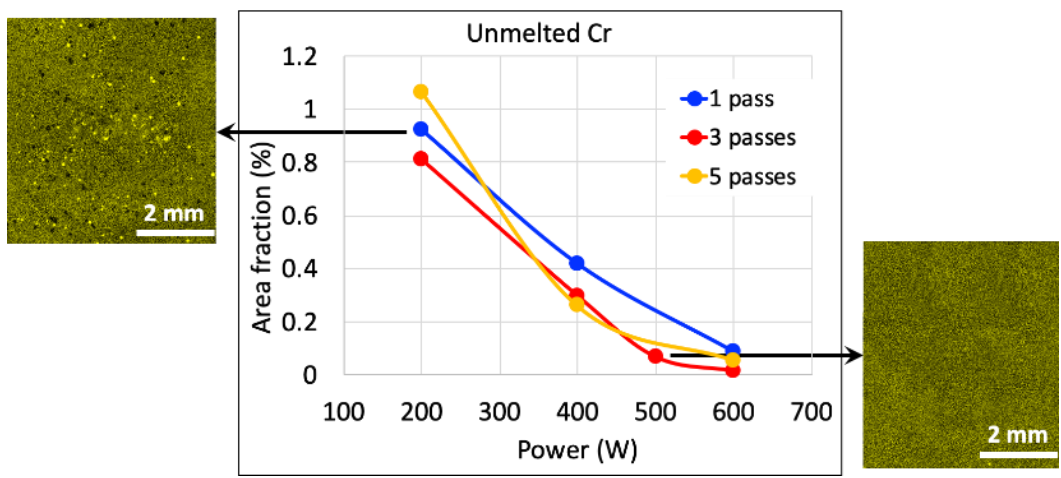

**Figure 2:** *Plot of the area fraction of unmelted Cr powders as a function of remelting parameters. Also shown are example EDS area maps at high and low unmelted Cr volume fractions.*

After optimization of the laser parameters to obtain good builds with minimum unmelted powders, it was necessary to calibrate the powder mass flow rates to achieve the right compositions in the alloys. The calibration process, first introduced in [42], was based on the relation between the elemental composition and the RPMs of the hopper augers:

$$x_i = \frac{R_i * M_i * [\alpha_i * (RPM)_i + \beta_i]}{\sum_{i=1}^{n} R_i * M_i * [\alpha_i * (RPM)_i + \beta_i]} \tag{1}$$

where, $x_i$ is the atomic fraction of element $i$ in the final build, $M_i$ is the molar mass of element $i$, $\alpha_i$ and $\beta_i$ are coefficients that relate the powder mass flow rates to the hopper RPM, and $R_i$ is the retention factor for each element $i$.

Multiple iterations of alloy printing were performed. After each iteration of 25 alloys, the samples were leveled, their compositions were measured via automated EDS and compared to the predicted compositions from equation (1). Any deviations from the predicted compositions were then corrected by varying $R_i$, $\alpha_i$, and $\beta_i$ as fitting parameters to obtain the least value for the sum of the residuals between the predicted and actual compositions. New auger RPMs were calculated based on the updated values of $R_i$, $\alpha_i$, and $\beta_i$, and were used to build the next iteration of alloys. This process was repeated over two more iterations until the deviation between the actual and predicted compositions reduced to less than 5 at%.

Going forward, all samples were built with the optimized print and remelt parameters (shown in Table 1). The alloy samples for indentation testing were printed in 5×5 arrays (25 samples) on 316L stainless steel build plates ($100 \times 100 \times 6.35$ mm). The nominal dimensions of each alloy sample were $10 \times 10 \times 5$ mm, with a spacing of 6.35 mm between samples. During printing, the laser hatch pattern was rotated 90° for every subsequent layer, regardless of whether it was the print or remelt layer.

**Table 1:** *The optimized laser parameters for the print and remelt layers. Note: All samples were manufactured using three remelt layers between every print layer.*

|  | Power (W) | Scan speed (mm/s) | Hatch spacing (mm) | Layer thickness (mm) |
|---|---|---|---|---|
| Print layer | 350 | 10.6 | 0.381 | 0.19 |
| Remelt layer | 500 | 21.2 | 0.381 | 0.19 |

## 2.2. Post-processing

After fabrication, the samples were heat treated in an ultra-high vacuum furnace with tungsten heating elements (manufactured by Materials Research Furnaces, LLC), that was able to achieve a vacuum of ~ $6.9 \times 10^{-2}$ Pa (~$10^{-5}$ psi). The heat treatment involved homogenization at 1000 °C for 24 hours. Two plates with the samples still attached were heat treated in a single run, allowing the simultaneous homogenization of 50 alloys.

After heat treatment, the samples were leveled to the same height via wire electrical discharge machining (EDM). After this, metallographic preparation involved grinding and polishing the samples while they were still attached to the plate, enabling simultaneous sample preparation of multiple samples. Grinding was performed on sequentially finer grit-papers, starting with 320-grit and ending on 1200-grit paper. The samples were then polished with diamond suspension in a glycol-based solvent, starting with 6-micron and ending on 1-micron. Final polishing was performed using 0.05-micron colloidal silica.

## 2.3. Characterization

All samples were characterized by energy-dispersive X-ray spectroscopy (EDS) on a JEOL JSM-6610 scanning electron microscope (SEM). EDS line scans were used to measure the chemical compositions of the samples, while EDS area maps were used to quantify the unmelted powder area fraction.

X-ray diffraction (XRD) was performed on a Bruker D8 Discover system to identify the phases and their crystal structures. Each sample on the build plate was assigned unique (x, y) positions. The sample stage was then programmed to move to the (x, y) location of each sample and perform the X-ray scan, before moving on to the next sample. This automated XRD acquisition eliminated the need to load and focus individual samples as is done conventionally. A single plate of 25 samples was completed in 100 minutes, without the need for user presence during the acquisition. Each sample was scanned at eight 2Θ positions – 20°, 32°, 44°, 56°, 68°, 80°, 92°, and 104° – for 30 s each.

The calculated phase diagrams (CALPHAD) method was utilized to predict the equilibrium phases at 1000 °C in the Cr-Fe-Mn-Ni alloy system. The calculations were performed using Pandat software (developed by Computherm, LLC) equipped with the PanHEA thermodynamic database.

Vickers hardness of all samples was measured on a Buehler Micromet II digital microhardness tester, with a pyramidal diamond indenter. High-throughput nanoindentation experiments were carried out on a Hysitron TI 950 TriboIndenter with a Berkovich tip. Each alloy sample was indented 15 times in the load-controlled mode with a loading rate of 1600 µN/s. The peak load was increased linearly, from 3000 µN for the first indent to 4000 µN for the fifteenth indent. The indents were performed in a 5x3 array with 15 µN spacing between the rows and columns to avoid interactions between their strain fields. Elastic modulus and hardness were determined from the indentation load-displacement curves using the Oliver-Pharr method [47].

Heat treatment, SEM analysis, XRD analysis, Vickers hardness measurements, and nanoindentation experiments were performed while the samples were still attached to the build plate, enabling high-throughput characterization.

## 3. RESULTS AND DISCUSSION

### 3.1. High-throughput synthesis and characterization

Initially, a total of 115 alloy samples were synthesized via DED-based additive manufacturing. An example 10 cm x 10 cm build plate demonstrating 25 of the samples is shown in *Figure 3*. Each hopper contained an elemental powder, and composition targets for each sample were achieved by calibrating the flow and capture of each element during in situ alloying. Calibration and sample processing were completed within one week. The nominal composition range encompassed Ni-rich, Fe-rich, and near-equimolar (center) regions of the Cr-Fe-Mn-Ni quaternary composition space. The set of alloys also included unary (elemental Cr, Fe, and Ni), binary (CrFe, CrMn, CrNi, FeMn, FeNi, and MnNi), and ternary (CrFeMn, CrFeNi, CrMnNi and FeMnNi) compositions. The bulk printed samples exhibited structural integrity and reasonable dimensional control. For all subsequent results, the samples remained affixed to the base plate and all data collected from the polished surface normal to the build direction.

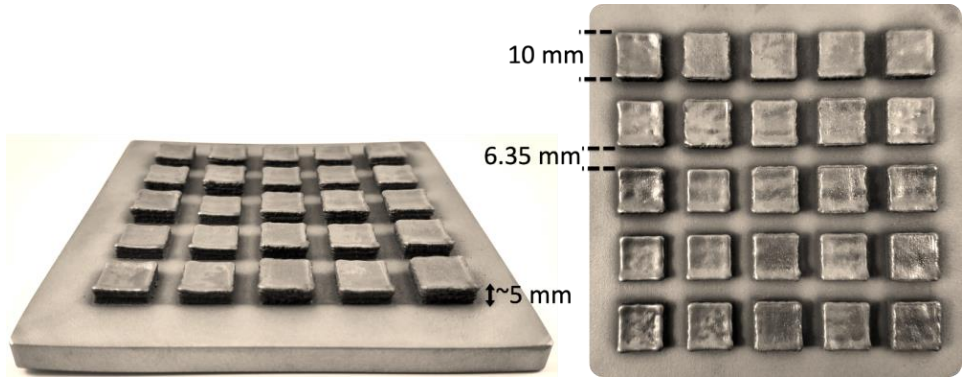

**Figure 3:** *A sample plate showing good dimensional consistency of the 3D printed samples printed approximately 5 mm in the build direction.*

All samples exhibited good chemical homogeneity and minimal porosity, comparable to that of traditional wrought or cast materials. The as-built microstructures of three example alloys exhibiting different phase distributions are compared in Figure 4. The SEM images of a single-phase face-centered cubic (FCC) alloy at magnifications of 100X and 1000X are shown in Figure 4a and 4b, respectively. The grain boundaries are clearly visible, and within the grains there are no other discernible features. The SEM images of a multi-phase alloy with face-centered- and body-centered-cubic phases (FCC+BCC) at magnifications of 100X and 5000X are shown in Figure 4c and 4d, respectively. The dendritic microstructure exhibited a fine primary dendrite arm spacing of ~1-2 µm. The SEM images of a multi-phase alloy with BCC and sigma phases (BCC+σ) at magnifications of 500X and 50000X are shown in Figure 4e and 4f, respectively. The grains in this alloy were equiaxed and finer compared to those in the single-phase FCC and FCC+BCC microstructures. In both multi-phase alloys, the secondary phases precipitated as particles in the size range 200-500 nm. These particles were pulled out from the microstructure during polishing, as evidenced by the pit-like features surrounded by wavy scratch patterns in Figure 4d and 4f.

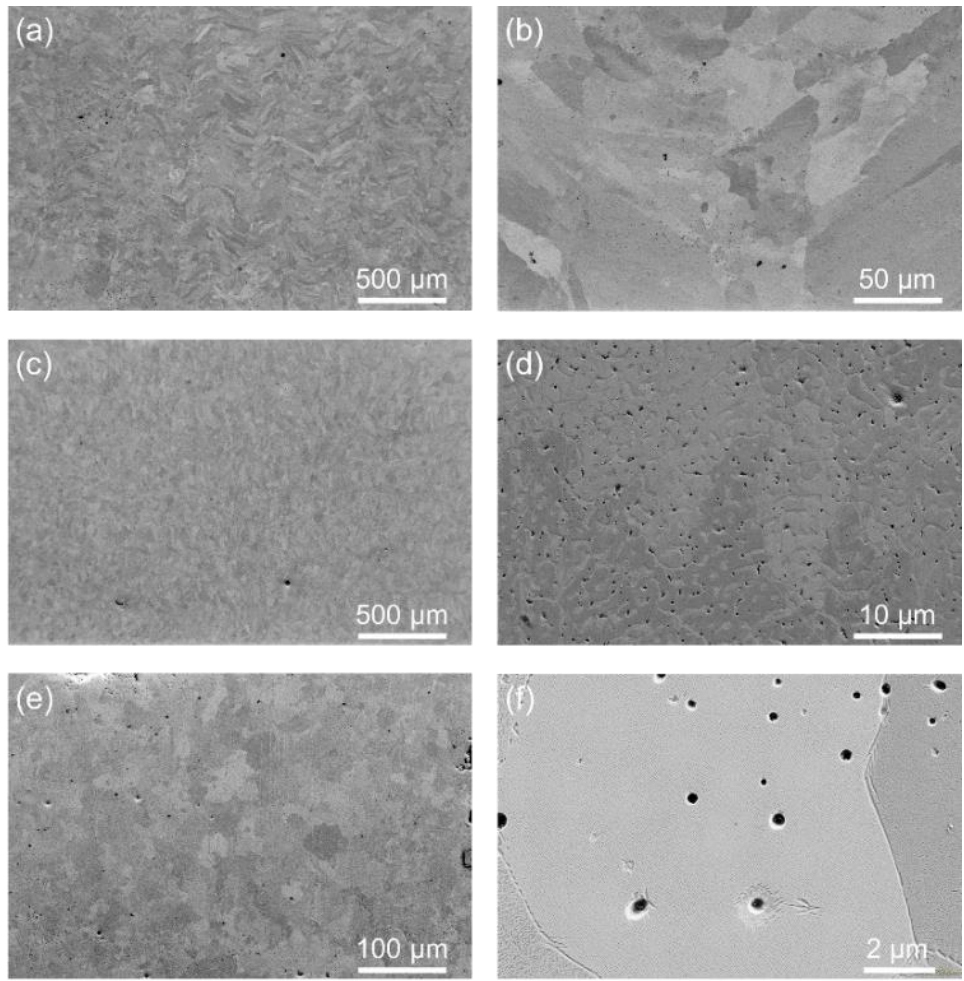

**Figure 4:** *Example microstructures in the as-built condition of (a) and (b) single-phase FCC alloy, (c) and (d) FCC+BCC alloy, and (e) and (f) BCC+σ alloy.*

After synthesis, all 115 alloys underwent a homogenization heat treatment at 1000 °C in a vacuum environment of ~ $6.9\times10^{-2}$ Pa (~$10^{-5}$ psi) for 24 hours followed by furnace cooling. The thermal cycle reduced elemental micro-segregation and alleviated residual stresses. Post-heat treatment, the alloys were characterized for composition via EDS line scans and phases present via XRD. An example of rapidly measured XRD patterns of 25 alloys additively manufactured on a single build plate is shown in Figure 5b. Having an array of discrete bulk alloys on a single build plate eliminated the SEM venting-evacuation cycle times between samples, which is typical for conventional samples. The compositions and phases of the additively manufactured and heat-treated alloys are summarized in Figure 5a, which shows the tetrahedron of the Cr-Fe-Mn-Ni MPEA composition-space. The different colored polyhedra within the composition space represent the expected phase regimes at 1000 °C based on CALPHAD predictions using Pandat software and PanHEA thermodynamic database. FCC, BCC, σ, FCC+BCC, FCC+σ, BCC+σ, and cubic A13 (β-Mn) phases are clearly visible. The synthesized alloys are represented by the blue, green, red, and yellow circles representing FCC, FCC+BCC, BCC, and BCC+σ alloys, respectively. Most of the synthesized exhibited phases in agreement with the CALPHAD-based predictions, validating the PanHEA thermodynamic database. However, a few alloys differed from predictions, as evidenced by the red circles in the blue regime. These would provide valuable input to improve the respective thermodynamic databases and solidification models.

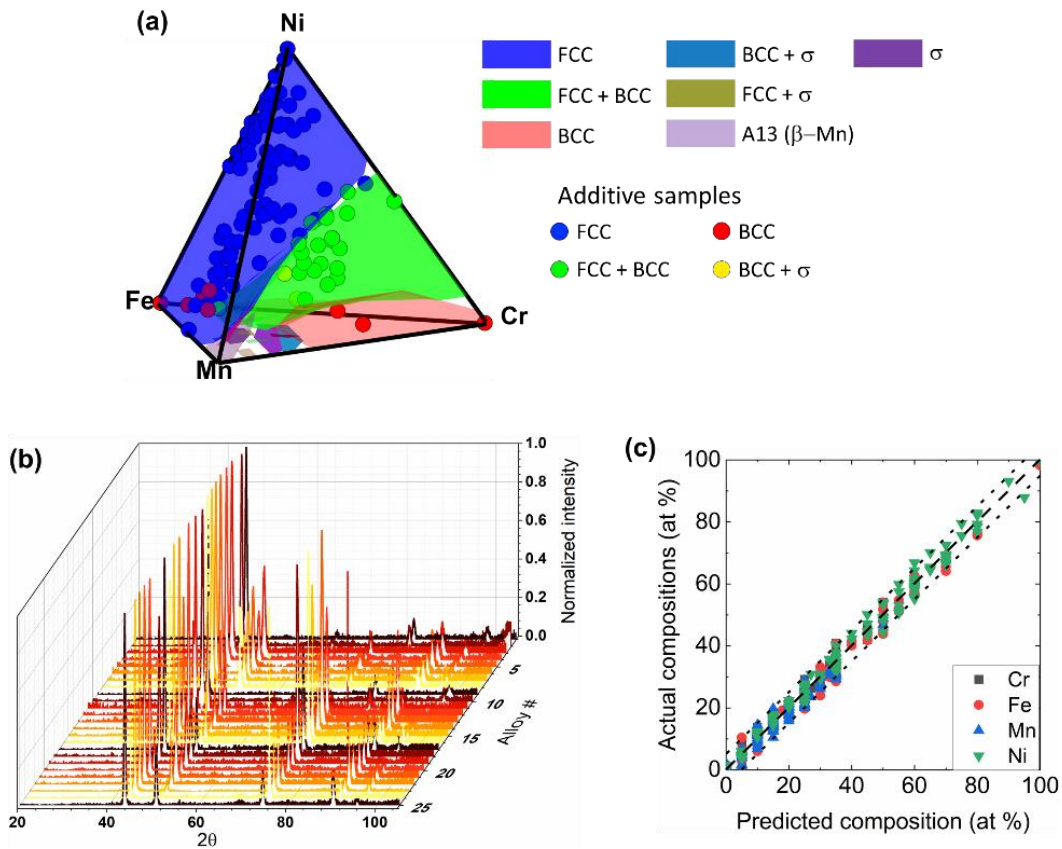

**Figure 5:** *a) The 4-element composition space. The alloys synthesized in this work are denoted by the different colored circles. The different colored polyhedra represent different phase regimes at 1000 °C as predicted by CALPHAD; b) Example XRD patterns of 25 3D-printed alloys obtained from automated analysis of a sample-plate; c) Plot showing actual vs. predicted compositions for the additively manufactured MPEA samples.*

The calibrated DED process allowed reasonable agreement between the predicted and actual compositions, shown graphically in Figure 5c. All actual element compositions were within an error of ±5 at% of the nominal (intended) composition, as indicated by the dotted lines in Figure 5c. The introduction of remelt laser passes after every print layer and strict calibration of the mass flow rates enabled tighter compositional control in the final alloys.

### 3.2. Mechanical properties

Going forward, it was decided to focus only on FCC alloys because of their potential for diverse applications that require ductility, toughness, and creep resistance. FCC alloys also offered a superior training set to develop a solid solution strengthening model without the complexities of secondary phases. Of the 115 additively manufactured and heat-treated alloys, 83 were single-phase FCC with unique compositions. To evaluate the mechanical properties of the homogenized alloys, Vickers hardness tests were performed on all samples without separating them from the build plate, allowing for quick evaluation of their mechanical strength. Nanoindentation experiments were also performed on these alloys in an automated fashion to enable rapid measurement of nanohardness and Young's modulus. The micro- and nanohardness trends were the same, with the microhardness representing a larger volume of tested material. The microhardness and Young's modulus values measured for single phase FCC alloys are depicted in Figure 6a and 6b, respectively. The alloys exhibited a wide range of hardness values, from ~ 140 HV0.05 to 280 HV0.05. The Young's modulus values ranged from ~190 GPa to 250 GPa.

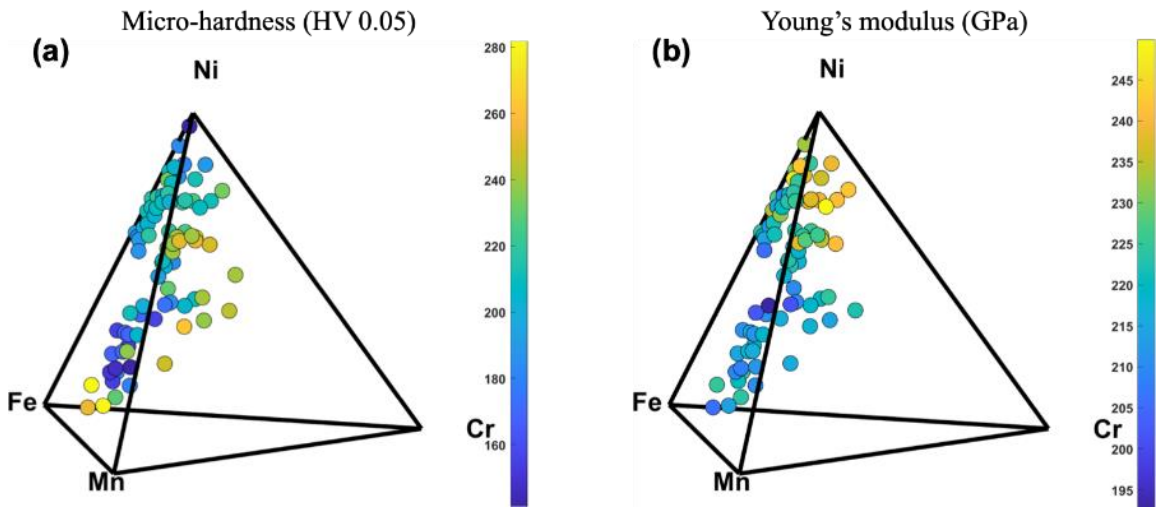

**Figure 6:** *3-D composition tetrahedra representing the (a) microhardness and (b) Young's modulus values of all FCC alloys.*

The high-throughput synthesis and characterization techniques developed in this work enabled visualization of a broad, quaternary composition-space and allowed a comparison of sample properties. The Fe-rich alloys exhibited the highest microhardness values, but low modulus values. On the other hand, Ni-rich alloys exhibited low microhardness but high modulus values. Alloys in the center of the composition-space (medium- and high-entropy alloys) exhibited a combination of increased microhardness and Young's modulus.

### 3.3. Machine learning

The ability to rapidly produce alloys of varying compositions and measure their hardness enabled the building of predictive computational models. However, when exploring new materials and synthesis techniques, the data generated in materials science can often be noisy due to the inherent complexity of the process. This can result in recorded features and measured properties that may not accurately represent the true behavior of the material. Furthermore, human bias can lead to false positives being pursued during the iterative process of materials development and optimization, which can prevent the discovery of true global maxima.

One advantage of machine learning models is that, with proper training, they can effectively distinguish between signal and noise in the data, average out outliers, and provide unbiased recommendations for which experiments to run next. By leveraging the power of machine learning, researchers can avoid wasting resources on false leads and focus on identifying the most promising avenues for materials development. In addition, using a multi-lens approach, where different model architectures are trained on the same dataset, can serve two purposes. Firstly, it can provide greater confidence that the underlying signal is present despite the noise in the data. Secondly, it can reveal any untracked run-to-run variance that may exist in the dataset.

To demonstrate this approach, we applied Random Forest Regression (RFR) and Gaussian Process Regression (GPR) models to predict the micro-hardness of new Cr-Fe-Mn-Ni alloys to be synthesized via DED. We then used a novel acquisition policy that combined several criteria to choose which alloys to prioritize for testing. Specifically, we prioritized alloys that (1) maximized the model's upper confidence bound, (2) were in regions where the predicted average micro-hardness had the maximum variance, and (3) were located some minimum distance away from a previously observed alloy. By using this multi-lens approach and acquisition policy, we were able to more effectively identify promising new alloys and accelerate the materials discovery process.

#### 3.3.1. *Building the machine learning model for single-phase alloys*

The initial dataset of 83 single-phase FCC alloys with unique compositions and their measured mechanical properties formed the training set to build the models. For each composition, an additional feature, deltaLP, was defined as the difference between experimentally measured and theoretical lattice parameters. The experimental lattice parameters were calculated from the XRD patterns by applying the Nelson-Riley regression on the lattice parameters calculated from each reflection within a pattern [48]. Theoretical lattice parameters were calculated using Vegard's law which is a rule-of-mixtures approximation based on the weighted average of the individual constituent elements in the solid solution.

The GPR model for micro-hardness used a kernel consisting of equally weighted Radial Basis Function and White Noise. The GPR model was trained on only the alloy compositions and had an $R^2$ of 0.607 across random 5-fold cross validation (CV).

For the RFR model, the alloy compositions were featurized using the standard Materials Agnostic Platform for Informatics and Exploration (Magpie) [43] feature set. Matminer, an open-source tool for materials data [49], was used to extract the compositional features from Magpie into a form suitable for machine learning. As a result, each composition was transformed into the default 132 Matminer composition-based features. Additionally, a GPR model was trained on the composition of the alloys to predict the deltaLP (described above). The predicted deltaLP was then used as an additional feature for the micro-hardness model. A baseline RFR model with all the Magpie features was hyperparameter tuned across 5-fold CV and yielded an $R^2$ of 0.607, nearly identical to that of the GPR model. Feature reduction was accomplished by dropping highly correlated features (Pearson coefficient > 0.9), resulting in a drop in the number of features from 132 to 14. The RFR model hyperparameters were re-tuned using 5-fold CV resulting in slightly degraded overall $R^2$ of 0.554.

The standard Scikit Learn Random Forest feature importance ranker was used [50] and deltaLP was on average among the top 50% of the features, along with a set of features related to magnetic moment, which are likely proxies for the proportion of Fe and Ni in the alloy. The term, deltaLP, represents the deviation of the lattice parameter from the ideal value and can be thought of as a measure of lattice distortion. The lattice distortion due to the atomic size mismatch of the different elements (and the resulting improvement in mechanical properties) has been proposed as one of the benefits of the high entropy of mixing in HEAs [51-53]. Mishra et al. [54] have distinguished between two types of lattice strain: one is caused by displacement of the atoms from their ideal FCC positions due to a size mismatch with their neighbors, and the other is distortion of the lattice around a dislocation core due to its strain field. In this work, the homogenized microstructures of the quaternary alloys exhibited distortions of the first type. Elements of different atomic radii were randomly distributed in the disordered FCC solid solution and resulted in local energy variations that dampened the free motion of dislocations, thus increasing the solid solution strength of MPEAs/HEAs [54].

The 5-fold CV scores for both the RFR and GPR models were nearly equivalent, but not very high. For instance, the variation of deltaLP as a function of composition is shown in *Figure 7*. The experimental data (black) showed substantially more scatter while the GPR predictions (blue) seemed to average out the experimental noise. Given the presence of experimental noise, the fact that RFR and GPR performed similarly in 5-fold CV, and that the 5-fold CV predictions of the RFR and GPR had an $R^2$ of 0.754 with one another, it stands to reason that both model architectures were identifying some signal from the data which could be used for model refinement and identifying new potentially interesting alloys.

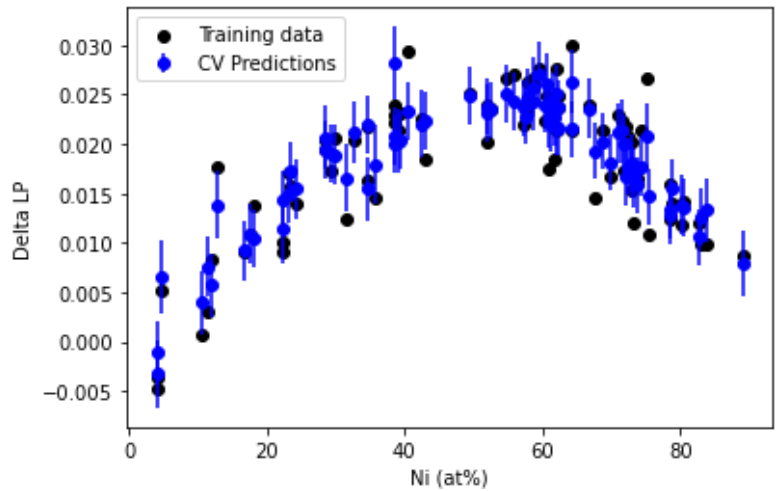

**Figure 7:** *Variation of delta lattice parameter as a function of Ni composition for the original experimental values (black) and the 5-fold CV predicted values (blue).*

For all subsequent predictions, an RFR model was trained on the full dataset and used to predict the micro-hardness values for 788 potential alloys in the Cr-Fe-Mn-Ni system. These alloys were calculated via CALPHAD methods to be FCC forming. The new alloys were featurized as described above and the RFR micro-hardness model was used to generate predictions for every alloy. The 95% confidence intervals for the predicted hardness values were calculated using quantile random forests [55] which queried the predicted values for each composition from each tree in the random forest. The predicted hardness values for the 788 alloys along with their 95% confidence bounds are shown in Figure 8a, ordered by their predicted mean hardness. The 95% confidence intervals ranged between -27.5 and 37.4 HV 0.05, and varied widely across the potential alloy composition space. For instance, hardness predictions exhibit substantially higher uncertainty for alloy numbers between 400 and 500, and for alloy numbers higher than ~720.

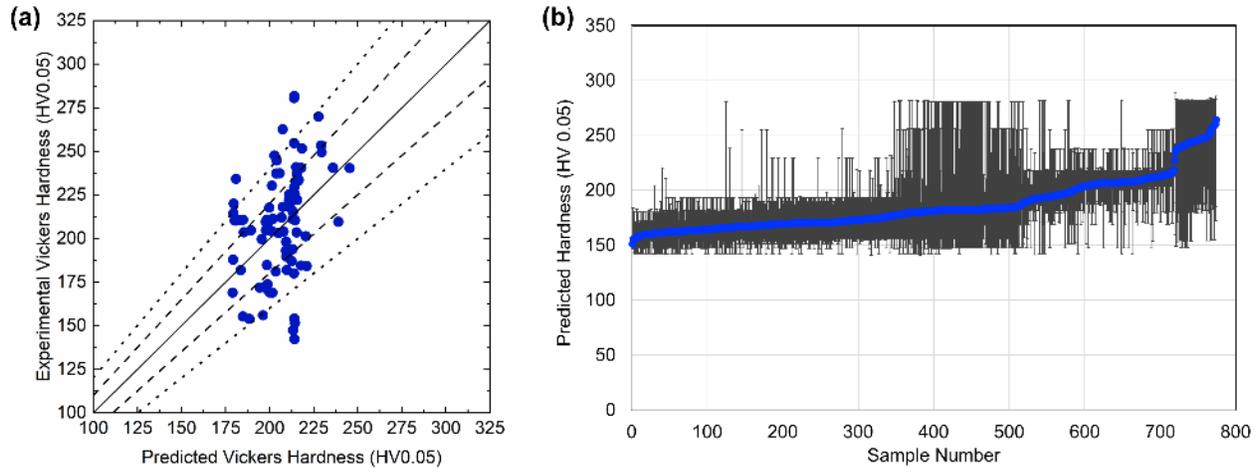

**Figure 8:** *a) Parity plots comparing model predictions with experimentally determined values of Vickers hardness. The solid line denotes a 1:1 parity, the dashed lines denote ±10% deviation from predictions, and the dotted lines denote ±20% deviation from predictions; b) Plot of the RFR model-predicted hardness values (blue) along with the 95% confidence intervals (dark grey) vs. sample number for the 788 hypothetical alloys.*

The 5-fold CV model output for the micro-hardness predictions is also displayed in the form of a parity plot of experimentally measured vs. predicted properties in ***Figure 8***b. Mean absolute percentage error (MAPE) was 12.1%, which is reasonable for a model built to be generalizable.

*3.3.2. Acquisition Policy and Material Selection*

To narrow down the number of alloy compositions recommended for the second AM iteration from 788 to 42, a novel acquisition policy was developed. First, the 788 potential alloys were clustered by their features into 40 clusters using k-means clustering [56]. The variability of each cluster, defined as the mean divided by the standard deviation, was calculated. The 20 clusters with the highest variability were kept, as these were places of either materials science interest or where the model is behaving poorly. Next the upper confidence bound was used to isolate the subset of alloys with the highest probability of having extraordinary properties or improving the overall model uncertainty. Finally, the Euclidean distance between the predicted and training points was calculated and all points within half of the average minimum distance between the datasets were dropped. Note that if the reader is rerunning the code, random seeds have explicitly not been set and run-to-run results may vary.

The resulting set of 42 new alloy compositions was additively manufactured, homogenized, leveled, and polished. Vickers hardness tests were performed to measure microhardness values. The results are plotted in Figure 9 which compares the microhardness values of alloys from the first iteration (Figure 9a) and the second iteration (Figure 9b).

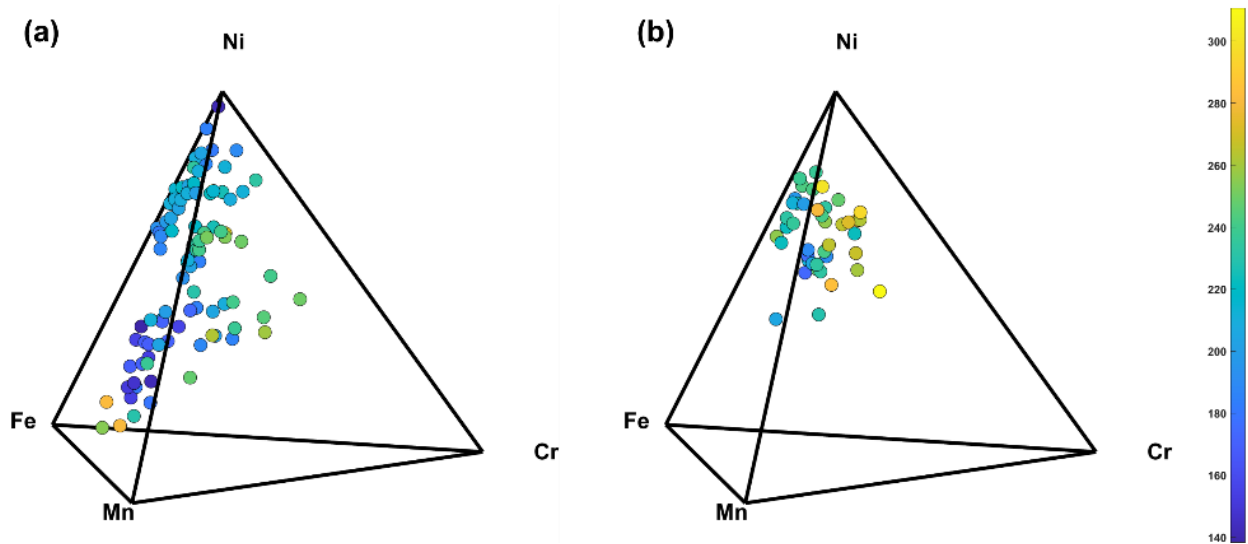

**Figure 9:** *The microhardness values of alloys from a) iteration-1 and b) iteration-2.*

Quantitative comparison of the results from the second LENS iteration and their predicted values showed an $R^2$ of -0.3, which was expected considering the new data points were selected to balance attacking overall model uncertainty and maximum expected improvement. The RFR and GPR model predictions showed strong correlation with each other (0.713), which indicated again that they picked up on similar signal. Incorporating the values from the second iteration of model training saw a 4.7% overall reduction in the prediction uncertainty across the entire 788 search space, indicating that the acquisition of about 5% of the predicted datapoints resulted in a proportionate reduction in the model uncertainty. Note that the entire test set was included here, as the experimental variation meant that none of these alloys was precisely present in the original test set.

In the first iteration of 83 FCC alloys, the hardness values ranged from 138.3 to 281.7 HV 0.05. The calculated average hardness was 207.7 ± 32.7 HV 0.05. In the second iteration of experiments of the 42 new FCC alloys, the hardness values ranged from 173.7 to 310.7 HV 0.05. The calculated average hardness was 239.4 ± 33.0 HV 0.05. Thus, it was evident that the active learning model successfully targeted new alloys that sampled from composition-spaces that were statistically harder. This is significant, as the most substantial errors from iteration-1 were in the region of high predicted hardness values, and as such, the active learning model sought to populate this region with additional data points to improve the model predictability. Based on this, four new alloys were identified that were harder than the hardest alloy from iteration-1. Their compositions and microhardness values are shown in Table 2.

**Table 2:** *Compositions (in at%) and micro-hardness values of four alloys from iteration-2 that exhibited the highest hardness.*

|         | Cr | Fe | Mn | Ni | Hardness (HV 0.05) |
|---------|----|----|----|----|---------------------|
| Alloy 1 | 33 | 23 | 1  | 43 | 310.7               |
| Alloy 2 | 5  | 10 | 10 | 75 | 300.2               |
| Alloy 3 | 20 | 15 | 0  | 65 | 295.9               |
| Alloy 4 | 21 | 33 | 2  | 44 | 284.1               |

The four alloys are also shown pictorially in the 3D composition tetrahedron in Figure 10a. XRD scans performed on these alloys (shown in Figure 10b) confirmed that the new "superhard" alloys were in fact single-phase FCC. Thus, the approach of balancing exploration and exploitation by targeting the upper confidence bounds in model predictions worked well in guiding us toward new materials that were harder than previously synthesized FCC alloys in this quaternary system.

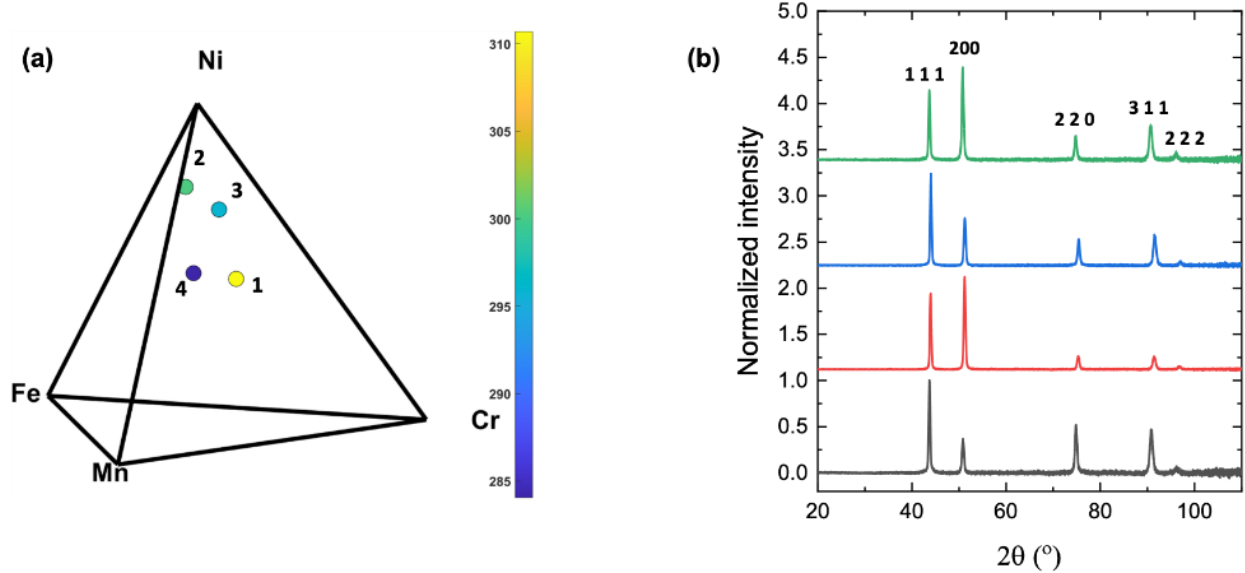

**Figure 10:** *a) 3D composition tetrahedron showing the four new alloys identified for their hardness, b) XRD patterns from these four alloys showing that they are all single-phase FCC.*

## 4. CONCLUSION

      DED-based additive manufacturing proved to be an effective high-throughput technique for synthesizing bulk discrete alloys, with orders of magnitude improvements in time savings compared to standard techniques such as arc melting. The current DED-based technique was able to produce 25 discrete bulk alloys (with sample dimensions ~ 10 mm × 10 mm × 5mm) in four hours. This translated to more than 100 alloys in a week (or 20 hours of equipment run time). The advantage of DED-based in situ alloying lies in the fact that it is inherently a parallel process. High-throughput synthesis is achieved by fabricating and evaluating multiple alloy samples simultaneously. On the other hand, conventional techniques like arc melting or induction melting are akin to series processes. Individual alloys of a single composition are synthesized, then characterized, and evaluated, before moving to a different alloy.

      From a reduced-order model perspective, the emergence of the parameter, deltaLP, was the prime predictor of microhardness and elastic modulus in these alloys, overshadowing all other features. The 5-fold LOCO CV method was more robust and conducive to extrapolation/generalization, albeit with higher errors. Combining this with an active learning approach is powerful because high-error regions can be selectively targeted in each successive iteration of experiments and continuously improve the model while maintaining generalizability.

Combining the experimental techniques with machine learning formed the basis for an iterative, closed-loop active learning framework that did not require additional datasets that could limit or misinform potential alloy design and selection strategies. In fact, the limitations of solely computation-based design strategies are their heavy reliance on accurate thermodynamic or property databases, as well as the ability to capture and represent all physical behavior in the computational models. The current self-contained active learning loop is not constrained in this manner. Moreover, with automation of the synthesis and characterization combined with in situ incorporation of the active learning models, autonomous systems can be envisioned.

## ACKNOWLEDGEMENTS

This work was supported by the Advanced Research Projects Agency-Energy (ARPA-E) program (ARPA-E DE-AAR0001050). The authors would like to thank Michael Niezgoda, graduate student in DJT's group at UW-Madison, for writing the g-code used to fabricate the DED samples in an automated fashion. The authors are also grateful to Prof. John H. Perepezko for access to his laboratory and the use of the high-temperature vacuum furnace. IS, VJ, and DJT gratefully acknowledge support from the US National Science Foundation through Designing Materials to Revolutionize and Engineer our Future (DMREF) award number 1728933.

## DATA AVAILABILITY

The data supporting the findings of this study are available upon request.

## REFERENCES

1. Greenfield, Aaron, and T. E. Graedel. "The omnivorous diet of modern technology." Resources, Conservation and Recycling 74 (2013): 1-7.

2. Davis, Joseph R. Aluminum and aluminum alloys. ASM international, 1993.

3. Agarwal, D. C. "Nickel and nickel alloys." Handbook of Advanced Materials (2004): 217.

4. Everhart, John, ed. "Engineering properties of nickel and nickel alloys." Springer Science & Business Media, 2012.

5. Reed, Roger C. The superalloys: fundamentals and applications. Cambridge university press, 2008.

6. Hume-Rothery, William. The structures of alloys of iron: an elementary introduction. Elsevier, 2016.

7. Bhadeshia, Harry, and Robert Honeycombe. Steels: microstructure and properties. Butterworth-Heinemann, 2017.

8. Motta, Arthur T., Adrien Couet, and Robert J. Comstock. "Corrosion of zirconium alloys used for nuclear fuel cladding." Annu. Rev. Mater. Res 45, no. 1 (2015): 311-343.

9. Charit, Indrajit. "Accident tolerant nuclear fuels and cladding materials." Jom 70, no. 2 (2018): 173-175.

10. Miracle DB, Senkov ON. A critical review of high entropy alloys and related concepts. Acta Materialia. 2017 Jan 1;122:448-511.

11. Cann, Jaclyn L., Anthony De Luca, David C. Dunand, David Dye, Daniel B. Miracle, Hyun Seok Oh, Elsa A. Olivetti et al. "Sustainability through alloy design: Challenges and opportunities." Progress in Materials Science 117 (2021): 100722.

12. Yeh, J-W., S-K. Chen, S-J. Lin, J-Y. Gan, T-S. Chin, T-T. Shun, C-H. Tsau, and S-Y. Chang. "Nanostructured high-entropy alloys with multiple principal elements: novel alloy design concepts and outcomes." Advanced Engineering Materials 6, no. 5 (2004): 299-303.

13. Cantor, Brain, I. T. H. Chang, P. Knight, and A. J. B. Vincent. "Microstructural development in equiatomic multicomponent alloys." Materials Science and Engineering: A 375 (2004): 213-218.

14. Gludovatz, Bernd, Anton Hohenwarter, Dhiraj Catoor, Edwin H. Chang, Easo P. George, and Robert O. Ritchie. "A fracture-resistant high-entropy alloy for cryogenic applications." Science 345, no. 6201 (2014): 1153-1158.

15. Li, Weidong, Peter K. Liaw, and Yanfei Gao. "Fracture resistance of high entropy alloys: A review." Intermetallics 99 (2018): 69-83.

16. Wang, Yafei, Bonita Goh, Phalgun Nelaturu, Thien Duong, Najlaa Hassan, Raphaelle David, Michael Moorehead et al. "Integrated High-Throughput and Machine Learning Methods to Accelerate Discovery of Molten Salt Corrosion-Resistant Alloys." Advanced Science (2022): 2200370.

17. Shi, Yunzhu, Bin Yang, and Peter K. Liaw. "Corrosion-resistant high-entropy alloys: a review." Metals 7, no. 2 (2017): 43.

18. Qiu, Y., S. Thomas, D. Fabijanic, A. J. Barlow, H. L. Fraser, and Nick Birbilis. "Microstructural evolution, electrochemical and corrosion properties of AlxCoCrFeNiTiy high entropy alloys." Materials & Design 170 (2019): 107698.

19. Otto, Frederik, A. Dlouhý, Ch Somsen, Hongbin Bei, G. Eggeler, and Easo P. George. "The influences of temperature and microstructure on the tensile properties of a CoCrFeMnNi high-entropy alloy." Acta Materialia 61, no. 15 (2013): 5743-5755.

20. Lu, Yiping, Yong Dong, Sheng Guo, Li Jiang, Huijun Kang, Tongmin Wang, Bin Wen et al. "A promising new class of high-temperature alloys: eutectic high-entropy alloys." Scientific reports 4 (2014): 6200.

21. Chen, PeiYong, Chanho Lee, Shao-Yu Wang, Mohsen Seifi, John J. Lewandowski, Karin A. Dahmen, HaoLing Jia et al. "Fatigue behavior of high-entropy alloys: A review." Science China Technological Sciences 61, no. 2 (2018): 168-178.

22. Shukla, Shivakant, Tianhao Wang, Shomari Cotton, and Rajiv S. Mishra. "Hierarchical microstructure for improved fatigue properties in a eutectic high entropy alloy." Scripta Materialia 156 (2018): 105-109.

23. Parkin, Calvin, Michael Moorehead, Mohamed Elbakhshwan, Jing Hu, Wei-Ying Chen, Meimei Li, Lingfeng He, Kumar Sridharan, and Adrien Couet. "In situ microstructural evolution in face-centered and body-centered cubic complex concentrated solid-solution alloys under heavy ion irradiation." Acta Materialia 198 (2020): 85-99.

24. Elbakhshwan, Mohamed, William Doniger, Cody Falconer, Michael Moorehead, Calvin Parkin, Chuan Zhang, Kumar Sridharan, and Adrien Couet. "Corrosion and thermal stability of CrMnFeNi high entropy alloy in molten FLiBe salt." Scientific reports 9, no. 1 (2019): 1-10.

25. Yang, Tengfei, Congyi Li, Steven J. Zinkle, Shijun Zhao, Hongbin Bei, and Yanwen Zhang. "Irradiation responses and defect behavior of single-phase concentrated solid solution alloys." Journal of Materials Research 33, no. 19 (2018): 3077-3091.

26. Kumar, NAP Kiran, C. Li, K. J. Leonard, H. Bei, and S. J. Zinkle. "Microstructural stability and mechanical behavior of FeNiMnCr high entropy alloy under ion irradiation." Acta Materialia 113 (2016): 230-244.

27. Yeh, Jien-Wei, Su-Jien Lin, Tsung-Shune Chin, Jon-Yiew Gan, Swe-Kai Chen, Tao-Tsung Shun, Chung-Huei Tsau, and Shou-Yi Chou. "Formation of simple crystal structures in Cu-Co-Ni-Cr-Al-Fe-Ti-V alloys with multiprincipal metallic elements." Metallurgical and Materials Transactions A 35, no. 8 (2004): 2533-2536.

28. Ranganathan, S. "Alloyed pleasures: multimetallic cocktails." Current science 85, no. 5 (2003): 1404-1406.

29. Miracle DB, Miller JD, Senkov ON, Woodward C, Uchic MD, Tiley J. Exploration and development of high entropy alloys for structural applications. Entropy. 2014 Jan;16(1):494-525.

30. Zhao, Ji-Cheng. "Combinatorial approaches as effective tools in the study of phase diagrams and composition–structure–property relationships." Progress in materials science 51, no. 5 (2006): 557-631.

31. Ludwig, Alfred, Robert Zarnetta, Sven Hamann, Alan Savan, and Sigurd Thienhaus. "Development of multifunctional thin films using high-throughput experimentation methods." International journal of materials research 99, no. 10 (2008): 1144-1149.

32. Gebhardt, Thomas, Denis Music, Tetsuya Takahashi, and Jochen M. Schneider. "Combinatorial thin film materials science: From alloy discovery and optimization to alloy design." Thin Solid Films 520, no. 17 (2012): 5491-5499.

33. Kube, Sebastian Alexander, Sungwoo Sohn, David Uhl, Amit Datye, Apurva Mehta, and Jan Schroers. "Phase selection motifs in High Entropy Alloys revealed through combinatorial methods: Large atomic size difference favors BCC over FCC." Acta Materialia 166 (2019): 677-686.

34. Kube, Sebastian A., and Jan Schroers. "Metastability in high entropy alloys." Scripta Materialia 186 (2020): 392-400.

35. Hofmann, Douglas C., Joanna Kolodziejska, Scott Roberts, Richard Otis, Robert Peter Dillon, Jong-Ook Suh, Zi-Kui Liu, and John-Paul Borgonia. "Compositionally graded metals: A new frontier of additive manufacturing." Journal of Materials Research 29, no. 17 (2014): 1899-1910.

36. Tsai, Peter, and Katharine M. Flores. "High-throughput discovery and characterization of multicomponent bulk metallic glass alloys." Acta Materialia 120 (2016): 426-434.

37. Gwalani, Bharat, Vishal Soni, Owais Ahmed Waseem, Srinivas Aditya Mantri, and Rajarshi Banerjee. "Laser additive manufacturing of compositionally graded AlCrFeMoVx (x=0 to 1) high-entropy alloy system." Optics & Laser Technology 113 (2019): 330-337.

38. Hochanadel, Patrick W., Robert D. Field, and Gary K. Lewis. "Microstructure and Properties of Laser Deposited and Wrought Alloy K-500 (UNS N05500)." Welding in the World 56, no. 11 (2012): 51-58.

39. Thoma, D. J., and J. H. Perepezko. "A Tri-Junction Diffusion Couple Analysis of the Nb-Cr-Ti System at 950 C." Experimental Methods of Phase Diagram Determination (1993): 45-54.

40. Zhao, J-C., Xuan Zheng, and David G. Cahill. "High-throughput diffusion multiples." Materials Today 8, no. 10 (2005): 28-37.

41. Wilson, Paul, Robert Field, and Michael Kaufman. "The use of diffusion multiples to examine the compositional dependence of phase stability and hardness of the Co-Cr-Fe-Mn-Ni high entropy alloy system." Intermetallics 75 (2016): 15-24.

42. Moorehead, Michael, Kaila Bertsch, Michael Niezgoda, Calvin Parkin, Mohamed Elbakhshwan, Kumar Sridharan, Chuan Zhang, Dan Thoma, and Adrien Couet. "High-throughput synthesis of Mo-Nb-Ta-W high-entropy alloys via additive manufacturing." Materials & Design 187 (2020): 108358.

43. Ward, Logan, Ankit Agrawal, Alok Choudhary, and Christopher Wolverton. "A general-purpose machine learning framework for predicting properties of inorganic materials." npj Computational Materials 2, no. 1 (2016): 1-7.

44. Ren, Fang, Logan Ward, Travis Williams, Kevin J. Laws, Christopher Wolverton, Jason Hattrick-Simpers, and Apurva Mehta. "Accelerated discovery of metallic glasses through iteration of machine learning and high-throughput experiments." Science advances 4, no. 4 (2018): eaaq1566.

45. Chang, Yao-Jen, Chia-Yung Jui, Wen-Jay Lee, and An-Chou Yeh. "Prediction of the composition and hardness of high-entropy alloys by machine learning." Jom 71, no. 10 (2019): 3433-3442.

46. Wen, Cheng, Yan Zhang, Changxin Wang, Dezhen Xue, Yang Bai, Stoichko Antonov, Lanhong Dai, Turab Lookman, and Yanjing Su. "Machine learning assisted design of high entropy alloys with desired property." Acta Materialia 170 (2019): 109-117.

47. Oliver, Warren Carl, and George Mathews Pharr. "An improved technique for determining hardness and elastic modulus using load and displacement sensing indentation experiments." Journal of materials research 7.6 (1992): 1564-1583.

48. Nelson, Jo Bo, and D. P. Riley. "An experimental investigation of extrapolation methods in the derivation of accurate unit-cell dimensions of crystals." Proceedings of the Physical Society 57, no. 3 (1945): 160.

49. Ward, Logan, Alexander Dunn, Alireza Faghaninia, Nils ER Zimmermann, Saurabh Bajaj, Qi Wang, Joseph Montoya et al. "Matminer: An open source toolkit for materials data mining." Computational Materials Science 152 (2018): 60-69.

50. Pedregosa, Fabian, Gaël Varoquaux, Alexandre Gramfort, Vincent Michel, Bertrand Thirion, Olivier Grisel, Mathieu Blondel et al. "Scikit-learn: Machine learning in Python." the Journal of machine Learning research 12 (2011): 2825-2830.

51. Yeh, Jien-Wei, Su-Jien Lin, Tsung-Shune Chin, Jon-Yiew Gan, Swe-Kai Chen, Tao-Tsung Shun, Chung-Huei Tsau, and Shou-Yi Chou. "Formation of simple crystal structures in Cu-Co-Ni-Cr-Al-Fe-Ti-V alloys with multiprincipal metallic elements." Metallurgical and Materials Transactions A 35, no. 8 (2004): 2533-2536.

52. Zhou, Y. J., Y. Zhang, Y. L. Wang, and G. L. Chen. "Solid solution alloys of Al Co Cr Fe Ni Ti x with excellent room-temperature mechanical properties." Applied physics letters 90, no. 18 (2007): 181904.

53. Zhang, Yong, Ting Ting Zuo, Zhi Tang, Michael C. Gao, Karin A. Dahmen, Peter K. Liaw, and Zhao Ping Lu. "Microstructures and properties of high-entropy alloys." Progress in materials science 61 (2014): 1-93.

54. Mishra, R. S., N. Kumar, and M. Komarasamy. "Lattice strain framework for plastic deformation in complex concentrated alloys including high entropy alloys." Materials Science and Technology 31, no. 10 (2015): 1259-1263.

55. Meinshausen, Nicolai, and Greg Ridgeway. "Quantile regression forests." Journal of machine learning research 7, no. 6 (2006).

56. Hartigan, John A., and Manchek A. Wong. "Algorithm AS 136: A k-means clustering algorithm." Journal of the royal statistical society. series c (applied statistics) 28, no. 1 (1979): 100-108.

57. Meredig, Bryce, Erin Antono, Carena Church, Maxwell Hutchinson, Julia Ling, Sean Paradiso, Ben Blaiszik et al. "Can machine learning identify the next high-temperature superconductor? Examining extrapolation performance for materials discovery." Molecular Systems Design & Engineering 3, no. 5 (2018): 819-825.

# Supplementary material

## 1. PROCESS CALIBRATION AND OPTIMIZATION

In situ alloying in the directed energy deposition (DED) process required three preliminary steps – measurement of the individual powder mass flow rates, optimization of the laser parameters for print and remelt layers, and composition calibration based on the previous two steps.

### 1.1. Powder mass flow rate calibration

In-situ alloying requires a known relationship between the RPM and the powder flow, and the first step was to measure the mass flow rates of each individual powder. For each hopper, the mass of the powder blown in one minute for a given RPM was measured. This was repeated for different RPM values for all the hoppers. The results are plotted in Figure 11. The linear equations generated from these measurements were the first step to predict the amount of powder required to achieve a target composition.

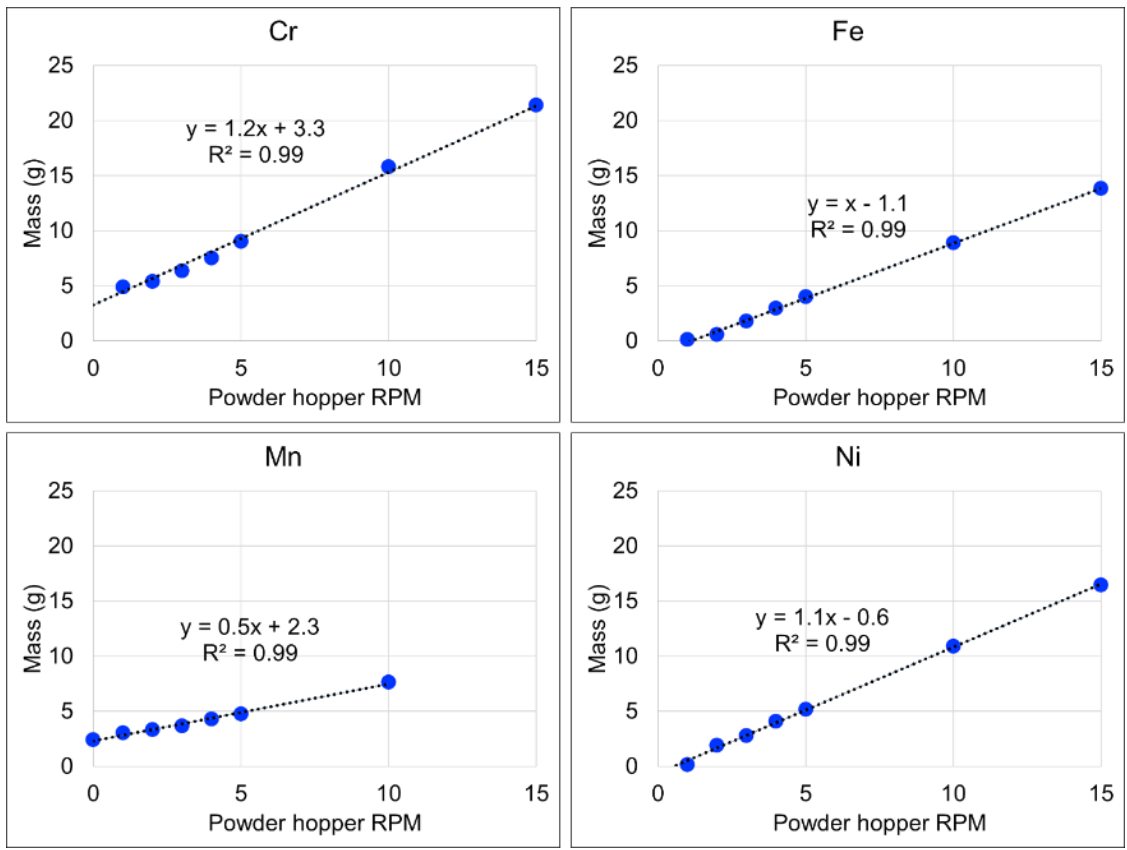

**Figure 11:** *Mass flow rate measurements of Cr, Fe, Mn, and Ni elemental powders.*

## 1.2. Laser parameter optimization

### 1.2.1. *Print parameter optimization:*

The laser parameters were optimized by building samples on two stainless steel build plates. 25 samples were built in a 5×5 matrix on each build plate, by varying the laser parameters while maintaining the predicted composition at an equimolar ratio. The first build plate, henceforth called the print plate (see Figure 12), was used to optimize the laser parameters of each print layer during the 3D printing process. The samples were printed by varying the laser power from 300 to 500 W in increments of 50 W for each row. The laser scan speed was varied from 10 to 30 in/min (4.2 to 12.7 mm/s) in increments of 5 in/min (2.1 mm/s) for each column (see Figure 12a). A single remelt pass was performed after each print layer with a laser power of 400 W and scan speed of 50 in/min (21.2 mm/s). The hatch spacing was 0.381 mm for the print layers and 0.19 mm for the remelting layers.

The heights of the sample increased steadily from the bottom-left to the top-right (as seen in Figure 12b), which also happened to be how the energy density increased. Samples built with the highest energy densities displayed excessive build and deformed shape. Laser power of 350 W and scan speed of 25 in/min (10.6 mm/s) were selected as the optimum print parameters based on which sample built closest to the target dimensions – 6.35×6.35 mm and height 5 mm (indicated by the yellow circle in Figure 12a). It is important to note that the "best" sample was chosen solely for its dimensional stability, and not based on microstructural analysis.

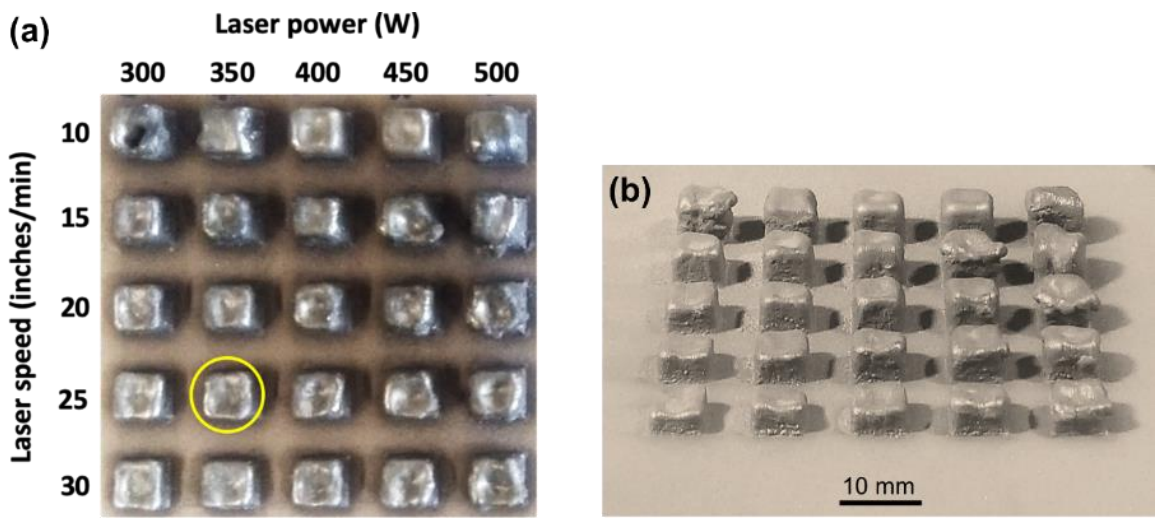

**Figure 12.** *a) Top-view of the print plate showing how the process parameters were varied during the building of the samples; b) side-view of the plate showing the variation of the dimensions of the samples.*

The melting point of Cr, the highest melting element in this alloy, is 1907 °C. The boiling point of Mn, the lowest melting element in the alloy, is 2061 °C. Because these two temperatures are close to each other, there is always an uncertainty of either boiling away the Mn due to the energy density being too high or retaining unmelted Cr powders in the built samples due to the energy density being too low.

To check for unmelted powders in the samples on the print plate, they were sliced via EDM so that all samples would be the same height. This enabled the simultaneous polishing of all the samples on the plate. Then, large-area EDS scans were performed, mapping for Cr, Fe, Mn, and Ni. On each alloy sample, six EDS area-scans were done with ~15-20% overlap. These six scans were then stitched together so that they represented a large area of ~3×4 mm of the sample surface (see Figure 13).

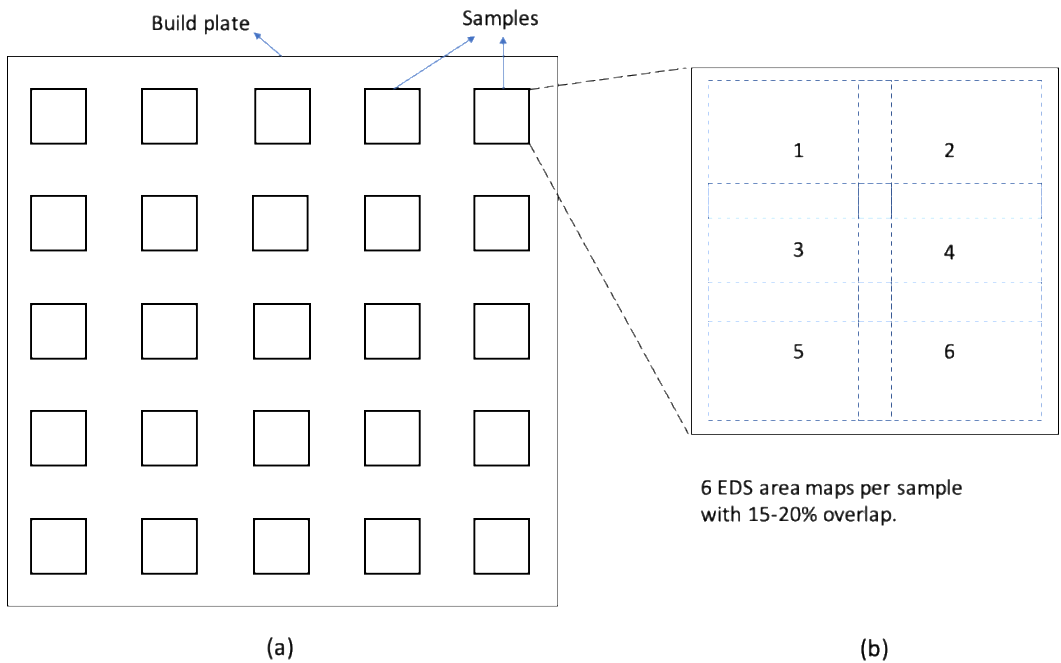

**Figure 13.** *a) schematic of additively manufactured build plate with a 5×5 array of samples, (b) schematic of a single sample with the dotted rectangles depicting the 6 EDS map areas with 15-20% overlap.*

All the samples exhibited unmelted powders, with Cr and Fe being the most problematic. As expected, there was a general trend of higher amount of unmelted elemental powders at the lower energy densities, and more complete melting at higher energy densities.

*1.2.2. Remelt parameter optimization:*

Since all the samples on the print plate exhibit unmelted elemental powders, there was a need to perform remelting passes between each printed layer of a sample. A remelt pass involved passing the laser over a layer that has just been printed, without any powder flowing. This would help in melting any powder that has remained unmelted in the current layer before printing the next layer. The use of remelt passes to eliminate unmelted powders has been previously reported by [1], however, they only used a single remelt pass on the top surface of a built sample.

The second build plate, henceforth called the remelt plate (Figure 14), was used to optimize the laser parameters for the remelting passes. The samples were printed by varying the laser power from 200 to 600 W in increments of 100 W for each row. The number of remelt passes between each print layer was varied from 1 to 5 for each column (see Figure 14). The laser scan speed was maintained at 50 in/min (21.2 mm/s). The printing parameters for each print layer were 350 W at 25 in/min (10.6 mm/s), based on the previous plate. The hatch spacing was 0.381 mm for the print layers and 0.19 mm for the remelting layers.

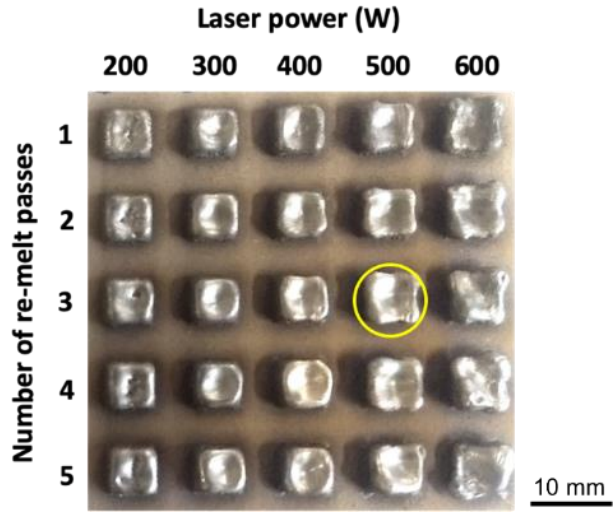

**Figure 14.** *Top-view of the remelt plate showing how the remelting parameters were varied.*

Like the previous plate, the samples were sliced via EDM and large-area EDS scans were performed to quantify the area fraction of the unmelted powders. The results are plotted in Figure 15 based on the EDS maps for Cr. A clear trend emerged where the area fraction of the unmelted Cr decreased rapidly with increasing laser power, almost dropping to zero at 500 and 600 W. On the other hand, the number of remelt passes had a lower effect on the melting of the powders.

There was only a slight decrease in the area fraction with increasing number of passes, at any given power. The single red data point represents the sample that was chosen as the best in terms of both, the least amount of unmelted powders as well as dimensional stability. This sample is also shown in the yellow circle in Figure 14.

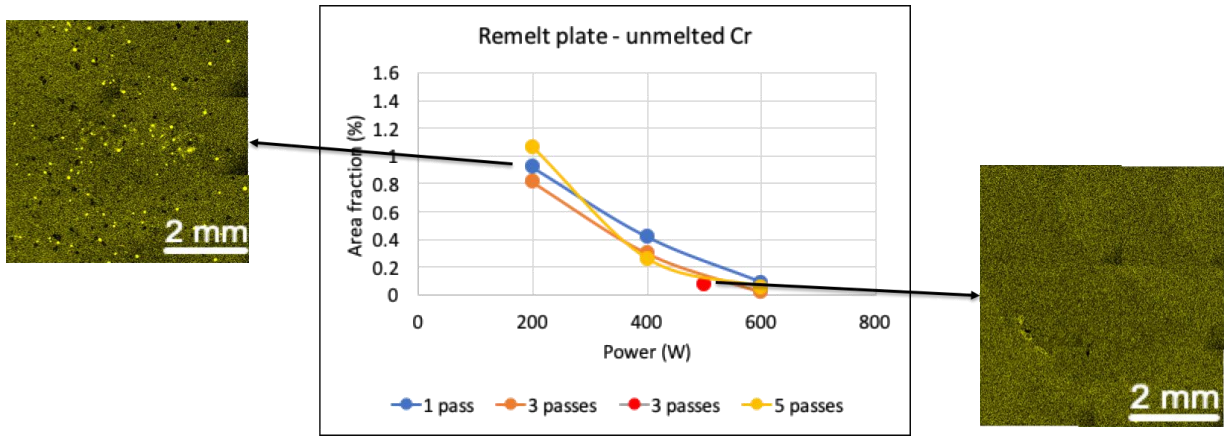

**Figure 15.** *Plot of the area fraction of unmelted Cr powders as a function of remelting parameters. Also shown are example EDS area maps at high and low unmelted Cr volume fractions.*

### 1.3. Composition calibration

In the previous stage, the composition of the samples was not varied, but kept constant at a predicted equimolar ratio. In this stage, the best laser parameters from both the print and remelt plates were used to fabricate multiple plates where the compositions of each sample were varied. The composition calibration technique outlined previously by [1] was followed.

Alloy samples of different compositions were built in a 5×5 matrix on a build plate. The compositions for this plate were centered around the equimolar composition, targeting the center of the HEA-composition space. The composition predictions were based on the linear equations obtained from the powder mass flow measurements described previously (see Figure 11 and related text). The linear equations were of the form [1]:

$$\dot{m}_i = \alpha_i * (RPM)_i + \beta_i \tag{1}$$

where, $\dot{m}_i$ is the mass flow rate of element *i, and* $\alpha_i$ and $\beta_i$ are coefficients that relate $\dot{m}_i$ to the hopper RPM. If $M_i$ is the molar mass of element *i*, then (1) can be rewritten to reflect the atomic fraction of element *i* as:

$$x_i = \frac{M_i * \dot{m}_i}{\sum_{i=1}^{n} M_i * \dot{m}_i} = \frac{M_i * [\alpha_i * (RPM)_i + \beta_i]}{\sum_{i=1}^{n} M_i * [\alpha_i * (RPM)_i + \beta_i]} \quad (2)$$

The compositions of the samples were measured via EDS line scans and compared to the predicted compositions based on equation (2). The was slight deviation between the actual and predicted compositions. To correct for this deviation a retention factor, $R_i$ (introduced in [1]) for each element $i$, was incorporated into equation (2), giving:

$$x_i = \frac{R_i * M_i * \dot{m}_i}{\sum_{i=1}^{n} R_i * M_i * \dot{m}_i} = \frac{R_i * M_i * [\alpha_i * (RPM)_i + \beta_i]}{\sum_{i=1}^{n} R_i * M_i * [\alpha_i * (RPM)_i + \beta_i]} \quad (3)$$

$R_i$, $\alpha_i$, and $\beta_i$ were varied as fitting parameters to obtain the least value for the sum of the residuals between the predicted and actual compositions. New auger RPMs were calculated based on the updated values of $R_i$, $\alpha_i$, and $\beta_i$, and were used to build the next iteration of alloys. This process was repeated over two more iterations until the deviation between the actual and predicted compositions reduced to less than 5 at%.

## 2. ALLOY COMPOSITIONS AND PROPERTIES USED FOR THE TRAINING SET

| Cr (at%) | Fe (at%) | Mn (at%) | Ni (at%) | deltaLP (Å) | Std. dev | Average hardness (HV 0.05) | Std. dev |
|---|---|---|---|---|---|---|---|
| 6.6 | 71.1 | 18.3 | 4.0 | -0.0035 | 0.0013 | 254.7 | 9.7 |
| 11.4 | 70.0 | 14.5 | 4.1 | -0.0048 | 0.0015 | 280.6 | 11.6 |
| 5.7 | 69.4 | 14.2 | 10.7 | 0.0006 | 0.0017 | 281.7 | 13.3 |
| 11.9 | 66.9 | 9.8 | 11.4 | 0.0030 | 0.0014 | 154.1 | 5.6 |
| 16.5 | 75.8 | 3.0 | 4.7 | 0.0052 | 0.0010 | 229.2 | 7.7 |
| 14.0 | 71.7 | 1.5 | 12.8 | 0.0176 | 0.0012 | 181.1 | 8.0 |
| 7.0 | 55.1 | 21.1 | 16.8 | 0.0091 | 0.0055 | 151.6 | 8.7 |
| 14.7 | 55.5 | 17.8 | 12.0 | 0.0084 | 0.0049 | 180.0 | 6.2 |
| 5.9 | 53.8 | 18.0 | 22.2 | 0.0102 | 0.0039 | 169.0 | 7.2 |
| 13.1 | 54.0 | 15.2 | 17.6 | 0.0108 | 0.0025 | 142.3 | 3.8 |
| 6.5 | 55.6 | 9.6 | 28.3 | 0.0203 | 0.0023 | 155.3 | 11.7 |
| 11.3 | 55.7 | 9.7 | 23.4 | 0.0157 | 0.0019 | 153.9 | 5.7 |
| 8.0 | 53.2 | 20.8 | 18.0 | 0.0138 | 0.0034 | 147.3 | 7.4 |
| 8.5 | 51.0 | 17.4 | 23.0 | 0.0172 | 0.0048 | 170.8 | 2.6 |
| 7.9 | 51.1 | 12.6 | 28.4 | 0.0195 | 0.0027 | 168.9 | 10.3 |
| 8.4 | 45.5 | 21.9 | 24.2 | 0.0139 | 0.0033 | 237.3 | 6.9 |
| 7.3 | 44.5 | 19.0 | 29.3 | 0.0173 | 0.0026 | 169.1 | 4.7 |
| 9.8 | 44.3 | 11.5 | 34.4 | 0.0218 | 0.0013 | 171.9 | 8.0 |
| 15.1 | 44.7 | 7.7 | 32.5 | 0.0205 | 0.0031 | 156.0 | 25.3 |
| 8.9 | 39.5 | 21.7 | 29.8 | 0.0206 | 0.0036 | 204.3 | 15.2 |
| 0.7 | 3.2 | 0.5 | 95.6 | 0.0037 | 0.0008 | 141.3 | 3.6 |
| 0.5 | 9.8 | 0.6 | 89.1 | 0.0087 | 0.0012 | 184.5 | 11.2 |
| 1.1 | 10.5 | 9.9 | 78.5 | 0.0125 | 0.0020 | 210.3 | 6.9 |
| 0.9 | 11.0 | 5.3 | 82.8 | 0.0119 | 0.0014 | 203.6 | 9.4 |
| 0.8 | 15.1 | 5.6 | 78.5 | 0.0159 | 0.0017 | 234.2 | 6.1 |
| 1.1 | 16.8 | 1.6 | 80.5 | 0.0142 | 0.0011 | 210.3 | 6.2 |
| 4.2 | 12.1 | 0.8 | 83.0 | 0.0099 | 0.0013 | 181.8 | 5.6 |
| 4.4 | 16.1 | 0.6 | 78.8 | 0.0141 | 0.0014 | 187.9 | 9.4 |
| 1.6 | 9.8 | 15.4 | 73.2 | 0.0120 | 0.0014 | 201.3 | 7.2 |
| 1.0 | 16.5 | 10.6 | 72.0 | 0.0217 | 0.0018 | 210.9 | 2.1 |
| 0.9 | 11.7 | 12.0 | 75.3 | 0.0108 | 0.0022 | 217.8 | 3.2 |
| 0.5 | 20.2 | 9.5 | 69.8 | 0.0166 | 0.0017 | 209.6 | 3.0 |
| 0.9 | 17.0 | 8.4 | 73.7 | 0.0165 | 0.0025 | 204.7 | 4.8 |
| 0.4 | 26.0 | 4.7 | 68.8 | 0.0213 | 0.0020 | 210.8 | 14.5 |
| 0.6 | 21.9 | 5.1 | 72.4 | 0.0208 | 0.0026 | 220.1 | 9.1 |
| 0.8 | 31.6 | 0.9 | 66.7 | 0.0238 | 0.0019 | 210.6 | 5.7 |
| 0.4 | 28.0 | 0.7 | 70.9 | 0.0229 | 0.0022 | 214.5 | 9.3 |
| 0.7 | 21.1 | 10.8 | 67.5 | 0.0145 | 0.0024 | 203.2 | 6.2 |
| 0.6 | 26.6 | 8.7 | 64.1 | 0.0214 | 0.0013 | 204.0 | 11.0 |
| 0.9 | 31.6 | 5.4 | 62.2 | 0.0248 | 0.0019 | 198.1 | 6.6 |
| 1.7 | 26.7 | 10.8 | 60.8 | 0.0174 | 0.0021 | 217.8 | 3.0 |
| 0.5 | 32.7 | 8.2 | 58.5 | 0.0266 | 0.0023 | 189.4 | 4.5 |
| 0.8 | 36.1 | 4.4 | 58.7 | 0.0263 | 0.0019 | 182.0 | 5.5 |
| 0.6 | 38.6 | 1.5 | 59.4 | 0.0277 | 0.0023 | 193.0 | 3.1 |
| 0.8 | 31.2 | 12.3 | 55.7 | 0.0270 | 0.0018 | 187.0 | 3.5 |
| 6.7 | 4.0 | 9.2 | 80.1 | 0.0119 | 0.0024 | 212.1 | 8.4 |
| 9.3 | 4.7 | 2.4 | 83.6 | 0.0099 | 0.0005 | 189.6 | 6.4 |
| 5.0 | 4.0 | 16.6 | 74.4 | 0.0215 | 0.0021 | 215.4 | 4.8 |
| 10.5 | 4.3 | 13.5 | 71.6 | 0.0173 | 0.0018 | 209.6 | 5.4 |
| 5.5 | 8.5 | 13.1 | 72.9 | 0.0202 | 0.0016 | 221.9 | 10.0 |
| 9.1 | 9.0 | 9.2 | 72.7 | 0.0154 | 0.0019 | 225.4 | 2.3 |
| 17.2 | 5.8 | 1.8 | 75.2 | 0.0267 | 0.0009 | 233.0 | 4.2 |
| 15.8 | 11.6 | 1.1 | 71.5 | 0.0223 | 0.0010 | 211.2 | 3.7 |
| 9.6 | 3.2 | 23.1 | 64.2 | 0.0300 | 0.0033 | 240.4 | 11.8 |
| 7.5 | 9.1 | 21.6 | 61.9 | 0.0276 | 0.0026 | 251.6 | 8.6 |
| 16.5 | 4.2 | 18.6 | 60.7 | 0.0248 | 0.0021 | 249.5 | 17.3 |
| 6.8 | 13.5 | 19.3 | 60.3 | 0.0224 | 0.0018 | 237.3 | 6.1 |
| 12.5 | 8.2 | 17.7 | 61.6 | 0.0236 | 0.0025 | 253.4 | 10.1 |
| 8.8 | 17.7 | 12.4 | 61.1 | 0.0223 | 0.0022 | 241.0 | 6.0 |
| 13.5 | 12.1 | 12.9 | 61.6 | 0.0184 | 0.0028 | 269.9 | 3.0 |
| 7.0 | 21.8 | 9.0 | 62.2 | 0.0237 | 0.0014 | 218.1 | 3.3 |
| 11.4 | 18.4 | 8.1 | 62.1 | 0.0214 | 0.0013 | 225.4 | 4.7 |
| 7.5 | 14.4 | 20.3 | 57.9 | 0.0262 | 0.0018 | 240.5 | 4.9 |
| 7.8 | 18.1 | 17.0 | 57.1 | 0.0219 | 0.0010 | 233.5 | 10.0 |
| 7.6 | 22.0 | 12.9 | 57.5 | 0.0235 | 0.0019 | 232.6 | 4.1 |
| 6.0 | 18.5 | 20.7 | 54.8 | 0.0266 | 0.0018 | 210.6 | 3.3 |
| 8.5 | 23.3 | 16.2 | 52.0 | 0.0239 | 0.0017 | 203.5 | 7.4 |
| 8.2 | 26.4 | 12.9 | 52.5 | 0.0235 | 0.0014 | 222.1 | 3.5 |
| 12.1 | 27.4 | 8.6 | 51.9 | 0.0203 | 0.0017 | 184.2 | 2.6 |
| 7.1 | 23.4 | 20.2 | 49.4 | 0.0250 | 0.0020 | 193.9 | 6.4 |
| 27.0 | 3.0 | 18.7 | 51.2 | 0.0329 | 0.0025 | 240.1 | 5.3 |
| 23.7 | 32.9 | 4.3 | 39.2 | 0.0213 | 0.0018 | 209.6 | 4.0 |
| 20.6 | 32.2 | 9.3 | 37.8 | 0.0221 | 0.0030 | 204.7 | 11.6 |
| 15.7 | 32.6 | 12.4 | 39.3 | 0.0231 | 0.0029 | 184.8 | 5.1 |
| 13.8 | 32.0 | 15.2 | 39.0 | 0.0240 | 0.0028 | 173.7 | 3.8 |
| 6.7 | 32.5 | 21.3 | 39.4 | 0.0294 | 0.0017 | 199.6 | 2.4 |
| 35.1 | 29.1 | 0.6 | 35.2 | 0.0146 | 0.0024 | 244.8 | 4.4 |
| 1.6 | 28.0 | 31.4 | 39.0 | 0.0229 | 0.0024 | 211.3 | 5.4 |
| 19.4 | 36.6 | 22.9 | 21.1 | 0.0092 | 0.0018 | 247.6 | 3.4 |
| 13.9 | 32.9 | 9.8 | 43.3 | 0.0184 | 0.0011 | 230.3 | 7.0 |
| 20.8 | 14.3 | 21.4 | 43.5 | 0.0226 | 0.0029 | 240.6 | 6.1 |
| 20.3 | 26.4 | 20.1 | 33.2 | 0.0123 | 0.0019 | 262.7 | 17.2 |
| 26.5 | 26.5 | 12.9 | 34.1 | 0.0163 | 0.0019 | 237.4 | 8.5 |

**REFERENCES**

1. Moorehead, Michael, Kaila Bertsch, Michael Niezgoda, Calvin Parkin, Mohamed Elbakhshwan, Kumar Sridharan, Chuan Zhang, Dan Thoma, and Adrien Couet. "High-throughput synthesis of Mo-Nb-Ta-W high-entropy alloys via additive manufacturing." Materials & Design 187 (2020): 108358.